\documentclass[aps,prb,twocolumn,floatfix,showpacs,nofootinbib]{revtex4}
\usepackage{graphicx}
\usepackage{amsmath}
\usepackage{amsfonts}
\usepackage{hyperref}
\usepackage{comment}

\long\def\symbolfootnote[#1]#2{\begingroup%
\def\thefootnote{\fnsymbol{footnote}}\footnote[#1]{#2}\endgroup}

\begin{document}

\title{Transient Dynamics of Confined Charges in Quantum Dots in the
  Sequential Tunnelling Regime}

\author{Eduardo Vaz and Jordan Kyriakidis}

\affiliation{Department of Physics, Dalhousie University, Halifax,
  Nova Scotia, Canada, B3H~3J5}

\begin{abstract}
  We investigate the time-dependent, coherent, and dissipative
  dynamics of bound particles in single multilevel quantum dots in the
  presence of sequential tunnelling transport.  We focus on the
  nonequilibrium regime where several channels are available for
  transport.  Through a fully microscopic and non-Markovian density
  matrix formalism we investigate transport-induced decoherence and
  relaxation of the system.  We validate our methodology by also
  investigating the Markov limit on our model.  We confirm that not only
  does this limit neglect the coherent oscillations between system
  states as expected, but also the rate at which the steady state is
  reached under this limit significantly differs from the non-Markovian
  results.  By a systematic analysis of the decay constants and
  frequencies of coherent oscillations for the off-diagonal elements
  of the reduced density matrix under various realistic tunneling rate
  anisotropies and energy configurations, we outline a criteria for
  extended decoherence times.
\end{abstract}

\date{\today}

% PACS: 73.23.-b Electronic transport in mesoscopic systems 73.23.Hk
% Coulomb blockade; single-electron tunneling 73.21.La Quantum dots

\maketitle

\section{Introduction}

It is a fundamental feature of quantum theory that the dynamics of an
isolated system will follow a unitary evolution, and thus be fully
reversible.  In practice, most quantum systems are influenced by
uncontrolled and inevitable interactions with an often incoherent
environment, and this influence will typically destroy this
deterministic evolution and result in the rapid vanishing of any
quantum coherence within the open system.~\cite{breuer_oqs}

A large body of theoretical work has been carried out throughout the
last half century which has pushed forward the investigation of the
dynamics of open quantum systems (OQS).  The seminal work by
Nakajima~\cite{Nakajima:1958p4987} and
Zwanzig~\cite{Zwanzig:1960p5087} on projection operator methods, as
well as the works by Kraus,~\cite{Kraus:1971p2939}
Lindblad,~\cite{Lindblad:1976p2837} and Gorini, Kossakowki, and
Sudarshan~\cite{Gorini:1976p2990} on quantum dynamical semigroups, and
by Davies~\cite{davies_1974} on Markovian master equations, have
rigorously established the conditions for positivity of physical
probabilities in open quantum systems.  Since then, Markovian
approaches have been successfully used~\cite{Suarez:1992p3725} to
study steady state phenomena on open systems where the past memory of
the system is neglected, or on closed unitary systems, where the
dynamics are deterministic and thus reversible.

The advent of technological applications for quantum coherence, such
as in quantum information processing and cryptography, as well as the
use of ultrafast laser pulse excitations, are requiring resolutions of
the quantum dynamics of a system down to a femtosecond
timescale.~\cite{shah_1999} This transient regime, which carries the
coherence and relaxation dynamics, cannot in general be described by a
coarse-grained Markovian limit,~\cite{markov_exeption} and a
non-Markovian approach is usually necessary.  On this point there has
been significant progress in recent years; the application of
non-Markovian theories to formal physical models in the area of
low-dimensional dynamical quantum systems has been extraordinarily
fruitful.  For example, electron transport by means of full counting
statistics,~\cite{Braggio:2006p244, Bagrets:2006p254} current and shot
noise analysis by diagrammatic techniques,~\cite{Thielmann:2005p166}
qubit dynamics in the presence of 1/f noise,~\cite{burkard:125317}
decoherence of qubit systems by means of effective Hamiltonians for
the reservoir,~\cite{Lee:2004p2953} electron spin dynamics in quantum
dots,~\cite{Coish:2004p4613} and decoherence in ballistic
nanostructures by projector operator
techniques,~\cite{Knezevic:2008p4596} and transient currents through
quantum dots by means of the reduced density operator
formalism.~\cite{moldoveanu_2009}

Formalisms based on non-equilibrium Green's functions (NEGF) have also
been successful in the analysis of much phenomena in mesoscopic
systems.~\cite{kadanoff_baym,haug_koch,haug_jauho} Notably, work in
both elastic and inelastic transport in quantum dots has been carried
out by means of NEGF or derivatives of this
formalism.~\cite{wacker_jauno_1998,wacker_jauno_1999} However, NEGF is
inherently a closed-system formalism where the system has Hamiltonian
dynamics,~\cite{datta_transport} and thus does not account for
irreversibility arising from interactions with an unseen, unknown, or
otherwise intractable environment.  Promising advances have been made
to extend NEGF to OQS, for example, by treating the environment as a
correction to the system's self-energy,~\cite{datta_2000} or by
calculating two-time correlation functions,~\cite{Knezevic:2003p3439}
effectively separating the time scales into transient and steady-state
regimes.  These and other approaches constitute significant progress
toward a full non-Markovian open system analysis within the NEGF
formalism.~\symbolfootnote[2]{For a rigorous compendium of the theory,
  developments, and general methods to approach OQS, see the excellent
  presentations by Alicki and Lendi,~\cite{Alicki:2007p4721} Gardiner
  and Zoller,~\cite{WGardiner:2004p5301} and Breuer and
  Pettrucione~\cite{breuer_oqs}.}

Even so, when the evolution of the many-body system states themselves
are sought, one must calculate the Green's functions and then extract
the density matrix, which in the non-equilibrium case may not be
unique.~\cite{Knezevic:2003p3439} An alternative formulation, one we
adopt in the present paper, is to work with the reduced density matrix
(RDM) directly.  This arguably provides a more intuitive description
of the dynamics of non-Markovian OQS, and it readily yields the actual
states of the system and quantum coherence between them.

Although the works cited above have covered a tremendous amount of
ground, to our knowledge there has not been a systematic analysis of
the effects of diverse energy and coupling regimes on the quantum
coherence in an OQS under the presence of transport-induced
relaxation.  Thus, the aim of this paper is to investigate the effects
of varying energy and coupling parameters on the non-Markovian
dynamics of a single few-electron open quantum dot under the RDM
formalism.  We focus on the sequential transport regime where two or
more channels are available for transport, and solve the RDM for the
system for the occupation probabilities and quantum coherence to
determine energy and coupling regimes leading to larger decoherence
times.

This paper is outlined as follows: In section~\ref{sec:evol-dens-matr}
we derive the evolution equations for the density matrix of the OQS in
the Born approximation, and a brief description of the Markov limit is
given for completeness.  Section~\ref{sec:results} develops the
non-Markovian transport theory for a single quantum dot nanostructure
coupled to semiinfinite leads.  The tunnelling Hamiltonian for the
transport model and the assumptions made for the system and reservoirs
are described in Sec.~\ref{sec:transport_model}.
Section~\ref{sec:kernel} presents in detail the derivation of the
analytical expressions for the transition matrix (the memory kernel)
containing the dynamics for this model.  In Sec.~\ref{sec:2_channels}
we illustrate the non-Markovian approach by considering the case of a
multilevel dot in a regime where two channels are available for
transport.  The analytical expressions for the elements of the RDM are
shown in Sec.~\ref{sec:pop_prob_2ch}, and results are compared with
the Markov limit under a variation coupling anisotropy to different
orbitals in the quantum dot.  The off-diagonal elements representing
the coherence in the system are presented in Sec.~\ref{sec:coherence},
as well as a thorough analysis of the effects of varying the dominant
energy and coupling parameters for this model.  A summary of the
results of the variations of the energy and coupling parameters is
given in Sec.~\ref{sec:summary}.  Finally, conclusions are presented
in Sec.~\ref{sec:conclusion}.

\section{Dynamics of the Open Quantum System}
\label{sec:evol-dens-matr}

We concentrate on the dynamics of an OQS which does not in general
follow a unitary evolution, and where an inherent irreversibility may
arise due to coupling with an essentially Markovian (memoryless)
environment.  When information is continually exchanged between an OQS
and an environment whose degrees of freedom are either unknown,
intractable, or uninteresting, the (possibly mixed) states of the
system alone are described not by state vectors, but rather by the RDM of
the system~\cite{blum}.  The RDM can be obtained by taking a partial
trace over the degrees of freedom of the environment on the total
density matrix for the OQS plus environment, such that
\begin{equation}
  \label{eq:rho_reduced}
  \rho(t) = Tr_{\text{env}}[\sigma(t)] = \sum_n W_n(t) |
  \psi_n(t)\rangle \langle \psi_n(t) |
\end{equation}
is the RDM of the system, $\sigma(t)$ is the total density matrix, and
$W_n(t)$ is the probability that the system state $|\psi_n(t)\rangle$
is occupied at time $t$.

For Eq.~\eqref{eq:rho_reduced} to represent a physical system, where the
diagonal elements describe the occupation probabilities and the
off-diagonal elements describe the coherence between the OQS states,
we require $\rho_{ii}(t)>0$, as well as $|\rho_{i,j}(t)|\leq 1$ for all
$i$ and $j$.\cite{rho_conditions}

The partial trace over the degrees of freedom of the environment has
the benefit of accounting for the influence of the
environment on the OQS, but it also has the drawback of only allowing
for limited or \emph{average} environment descriptions.
Although partial-trace-free approaches have been
developed~\cite{Knezevic:2002p4598}, to our knowledge they have only
addressed the steady state regime where memory effects have been
neglected.

Considering an environment with a much greater number of degrees of
freedom relative to those of the system, and for weak OQS-environment
coupling in a sequential tunneling regime (Born approximation), we can
assume that the system has negligible effect (no back-action) on the
environment, that the system is not correlated with the environment,
and that the environment is essentially in equilibrium at all
times.~\cite{blum} In this case, the total density matrix for the OQS
plus environment can be written as
\begin{equation}\label{eq:born_approx}
  \sigma(t) \approx \rho(t)\otimes \varrho(0).
\end{equation}
where $\varrho(0)$ is the environment's equilibrium density matrix.

The dynamics of the RDM in the Born approximation can be obtained by
an iterative integration of the Liouville-von~Neumann equation to
obtain the time dependence for higher order
processes,~\cite{breuer_oqs}
\begin{align}\label{eq:liouville_total}
  \nonumber \dot \rho(t) = & -\mathcal{L}(t)\rho(0) + \int_0^t \!
  dt'
  \mathcal{L}(t) \mathcal{L}(t') \rho(t'),\\
\end{align}
where $\mathcal{L}(t)\mathcal{O} = i \hbar^{-1}[H(t),\mathcal{O}]$ is
a Liouville superoperator for any operator $\mathcal{O}$
and Hamiltonian $H(t)$.

The matrix elements of the above generalized master equation for the
RDM in the interaction picture can be written in the form~\cite{blum}
\begin{equation}
  \label{eq:rho_dot_born}
  \dot \rho_{ab}(t) = \sum_{c,d} \int_0^t \! dt'
  \rho_{cd}(t') R_{abcd}(t - t') e^{i\gamma_{abcd} t'},
\end{equation}
where, $R_{abcd}(t - t')$ is the memory kernel characterizing the
system, the environment, and their coupling, and where $ \gamma_{abcd} =
\omega_{ab} - \omega_{cd}$, with $\omega_{ab}=(E_a - E_b) / \hbar$
denoting energy differences between OQS states.

The system of coupled integrodifferential convolution equations,
Eq.~\eqref{eq:rho_dot_born}, is solved by transforming it into an
algebraic system of equations in Laplace space,
\begin{equation}
  \label{eq:laplace_system1}
  \sum_{cd}\mathcal{W}_{abcd}(s)~\tilde{\rho}_{ab}(s)=\tilde{\rho}_{cd}(0).
\end{equation}
where,
\begin{align}
  \nonumber &\mathcal{W}_{abcd}(s)=\left[ \delta_{ca}\delta_{bd}\left(
      s+i\omega_{ab}\right) - \tilde{R}_{abcd}(s) \right] \\
  \nonumber \tilde{\rho}_{ab}&(t)\equiv \rho_{ab}(t)e^{-i\omega_{ab}t}, \ \
  \tilde{R}_{abcd}(t)\equiv R_{abcd}(t)e^{-i\omega_{ba}t}.
\end{align}
For a given set of basis states, an inverse Laplace transform of
Eqn.~\eqref{eq:laplace_system1} yields the time dependent solutions in
the Heisenberg picture.

Analytic solutions in Laplace-space can be obtained for only a few OQS
states ($\sim 2-4$), since the computational effort rapidly increases
with the number of available transport channels. For larger numbers of
channels, we adopt a numerical approach to solve the linear system in
Eq.~\eqref{eq:laplace_system1}.

In the regime of sequential tunnelling transport through a quantum dot
containing uncorrelated electrons, the important transport channels
are those single-particle system states $|\alpha \rangle$, whose
energy lies within the bias window.  These single-particle states
define the possible dynamical many-body states of the system as those
involved when the single particle states are empty or occupied.  Thus,
for $k$ channels, a minimum of $2^k$ many-body system states are
required (for empty or occupied).  Furthermore, $(2^k)^2$
density-matrix elements are required to describe the population
probabilities as well as the coherence between the states.  The
temporal evolution of the density-matrix elements themselves are
governed by $(2^k)^4$ transition tensor elements.  In practice,
symmetry considerations can reduce this number only by a constant
amount (for details see Ref.~[\onlinecite{blum}] pg. 292).

For the results in the following sections we perform the inverse
Laplace transform in the form of a Bromwich
integral~\cite{abramowitz+stegun} over a contour chosen such that all
singularities are to the left of the contour line. The integration is
numerically performed~\cite{integration} at each time step using a
combination of a 25-point Clenshaw-Curtis and a 15-point Gauss-Kronrod
integration method.

\subsection*{Markov Limit}

We validate the use of a non-Markovian approach by a brief comparison
with the long-time Markov limit.~\cite{blum} This approximation
assumes that the environment correlation functions vanish at such a
fast rate that the reversibility, and thus memory of the OQS is
essentially destroyed.~\cite{breuer_oqs} In such a case, $\dot
\rho(t)$ becomes local in time.
\begin{equation}
  \label{eq:rho_markov_1}
  \dot \rho_{ab}(t) \longrightarrow \sum_{c,d} \rho_{cd}(t)
  \int_{0}^{t} \! dt' R_{abcd}(t - t') 
  e^{i\gamma_{abcd} t'}.
\end{equation}
For time intervals $t-t'$ much greater than the environment's
correlation time $\tau$, the correlation functions for environment
operators rapidly become uncorrelated and decay to zero, $ \langle
F_k^\dag (t-t')F_{k'}\rangle \approx \langle F_k^\dag(t-t')\rangle
\langle F_{k'}\rangle \approx 0$.  Since the environment is
uncorrelated beyond this $t-t' \gg \tau$ time, there are no
contributions to the dynamics, and the upper integration limit
in~\eqref{eq:rho_markov_1} may be extended to infinity with negligible
error in the calculations.  The lower integration limit of
Eq.~\eqref{eq:rho_markov_1} can be taken to negative infinity, and the
memory kernel becomes a delta function of the channel energies.  This
leads to a Fermi's Golden rule for transitions with strict energy
conservation.  Thus, the coupled set of integro-differential evolution
equations become a coupled set of first order ordinary differential
equations,
\begin{equation}
  \label{eq:rho_markov_2}
  \dot \rho_{ab}(t) = \sum_{c,d} W_{abcd} ~\rho_{cd}(t) ,
\end{equation}
where $W_{abcd}$ is a time-local transition rate.

The Markov approximation is typically valid only in the long-time
limit and leads to a zero-average of the off-diagonal elements of the
OQS, thus the transition rates $W_{abcd}$ can be written as
$W_{aabb}\equiv W_{ab}$, since only transitions between diagonal
elements become relevant.

In the following sections we verify the validity of the Markov limit
only once the system has established a steady
state~\cite{Suarez:1992p3725}, and that both the \emph{approach}
and relaxation times to the steady state in this limit are unreliable.

\section{Dynamics in a Single Quantum Dot}
\label{sec:results}

\subsection{Transport Model}
\label{sec:transport_model}

\begin{figure}[tb!]
\begin{center}
  \includegraphics[width=20pc]{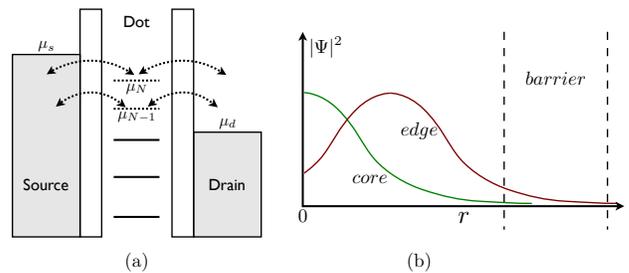}\hspace{2pc}%
  \begin{minipage}[b]{20pc}
    \caption{\label{fig:model}(color online) (a) Schematic
      representation of the single QD model coupled to source and
      drain reservoirs. Two transport channels are present within the
      bias window.  (b) Representation of a orbital asymmetry where
      edge ($p$-type) orbitals are more strongly coupled to the
      reservoir compared to core ($s$-type) orbitals.}
  \end{minipage}
\end{center}
\end{figure}

We consider a sequential transport model for a single quantum dot
coupled to semiinfinite leads as depicted in Fig.~\ref{fig:model}a.
The total Hamiltonian describing this coupled system plus environment
is given by,
\begin{subequations}
  \label{eq:model}
  \begin{equation}
    \label{eq:mainHamil}
    H = H_S + H_{QD} + H_D + \mathcal{H}_T,
  \end{equation}
  where $H_S$ and $H_D$ are the Hamiltonians describing the source and
  drain leads, respectively.  These are taken to be non-interacting
  Fermion systems shifted by the bias voltage $V_B$:
  \begin{equation}
    \label{eq:HReservoir}
    H_{S(D)} = \sum_{s(d)} (\epsilon_{s(d)} \pm \frac{1}{2} eV_B) 
    d^\dag_{s(d)} d_{s(d)},
  \end{equation}
  with $d^\dag_{s(d)}$ a creation operator for the source (drain), and
  $d_{s(d)}$ an annihilation operator.

  The Hamiltonian for the quantum dot in Eq.\eqref{eq:mainHamil} is
  given by,
  \begin{equation}
    \label{eq:QDotHamil}
    H_{QD} = \sum_i (\hbar\omega_i + eV_g) c^\dag_i c_i + V_{\mathrm{int}},
  \end{equation}
  where $c^\dag_i$ and $c_i$ are system creation and annihilation
  operators, the single-particle energies $\hbar \omega_i$ are all
  shifted by the applied gate voltage, $V_g$, and $V_{\mathrm{int}}$
  is the interaction among the confined particles.  In the present
  work we set $V_{\mathrm{int}}=0$.  Although we expect our
  qualitative results to remain unchanged, the implications of a full
  coulomb interaction in this transport theory may have significant
  consequences and will be the subject of future work.

  The tunneling Hamiltonian describing the coupling between the
  reservoirs and the dot is given by,~\cite{Mahan}
  \begin{equation}
    \label{eq:TunnelHamil}
    \mathcal{H}_T = \sum_{k, r = (s, d)} \left( T^r_k d^\dag_r c_k + h.c \right),
  \end{equation}
\end{subequations}
where $T^{s(d)}_k$ is a tunneling coefficient for a particle tunneling
from the single-particle state $|k\rangle$ in the dot to the source
(drain) reservoir.  For the present model, we assume a small level
broadening $\Gamma$, and low temperature such that $\Gamma \ll k_BT
\ll \Delta E$. In this case the reservoirs are in their respective
non-degenerate ground states.

\subsection{Memory Kernel}\label{sec:kernel}

In order to obtain the system dynamics for this transport model,
we derive the memory kernel $R_{abcd}(t)$ appearing in
Eq.~\eqref{eq:rho_dot_born}.  By evaluating the Liouville superoperator
appearing in Eq.~\eqref{eq:liouville_total} in the interaction
picture, and within the weak interaction limit to second order in
Eq.~\eqref{eq:TunnelHamil}, the time evolution of the RDM becomes,
\begin{align}\label{eq:liouville_eq_total}
  \nonumber \dot \rho(t) = -
  &\frac{i}{\hbar}Tr_{\mathcal{B}}[\mathcal{H}_T(t),\rho(0)\varrho(0)]\\
  -&\frac{1}{\hbar^2}\int_0^t dt'
  Tr_{\mathcal{B}}[\mathcal{H}_T(t),[\mathcal{H}_T(t'),\rho(t')\varrho(0)]].
\end{align}

The tunneling Hamiltonian \eqref{eq:TunnelHamil} can be rewritten in the
form,
\begin{equation}\label{eq:H_T}
  \mathcal{H}_T(t)=\sum_k \left(F_k(t)Q^\dag_k(t)+F^\dag_k(t)Q_k(t)\right),
\end{equation}
where $Q_i(t)=e^{iH_{QD}t}c_ie^{-iH_{QD}t}$, and where
\begin{equation}\label{eq:H_T_F}
  F_k(t)=\sum_sT^S_{k,s}L_s(t)+\sum_dT^D_{k,d}L_d(t)
\end{equation}
is a generalized reservoir operator that takes into account both
source and drain reservoir states, where
$L_{s(d)}(t)=e^{iH_{S(D)}t}d_{s(d)}e^{-iH_{S(D)}t}$.

Noting that the interaction Hamiltonian has zero expectation with
respect to the equilibrium ensemble of the environment, we have that
\begin{align}\label{eq:rho_dot_term1=0}
  Tr_R[F_k(t)\varrho(0)]=\langle F_k(t)\rangle=0,\\
  \nonumber \langle F_k(t)F_{k'}(t') \rangle =\langle F^\dag_k(t)F^\dag_{k'}(t')
  \rangle=0.
\end{align}

Thus, the odd moments of $\mathcal{H}_T$ as defined
in Eq.~\eqref{eq:H_T_F} with respect to the density matrix for the
equilibrium reservoir will all vanish.  

With Eq.~\eqref{eq:rho_dot_term1=0}, the evolution equation
(Eq.~\eqref{eq:liouville_eq_total}) becomes

\begin{align}\label{eq:rho_dot2}
  \nonumber
  \dot\rho(&t)=-\frac{1}{\hbar^2}\sum_{i,j}\int_0^tdt'\\
  \nonumber & \left\{
    \left(Q_i(t)Q^{\dag}_j(t')\rho(t')-Q^{\dag}_j(t')\rho(t')Q_i(t)\right)\langle
    F^{\dag}_i(t)F_j(t')\rangle
  \right.\\
  \nonumber
  +&\left(\rho(t')Q_i(t)Q^{\dag}_j(t')-Q_i(t)\rho(t')Q^{\dag}_j(t')\right)\langle
  F_i(t)F^{\dag}_j(t')\rangle\\
  \nonumber
  +&\left(\rho(t')Q_j(t')Q^{\dag}_i(t)-Q^{\dag}_i(t)\rho(t')Q_j(t')\right)\langle
  F^{\dag}_j(t')F_i(t)\rangle\\
  +&\left. \left(Q^{\dag}_i(t)Q_j(t')\rho(t')-Q_j(t')\rho(t')Q^{\dag}_i(t)
    \right)\langle F_i(t)F^{\dag}_j(t')\rangle \right\},
\end{align}

A matrix element of the first term on the right of \eqref{eq:rho_dot2}
(where there are 8 terms in total) can be rewritten in the following
way,
\begin{align}\label{eq:system_correlation}
  \nonumber \langle m'&|Q_i(t)Q^{\dag}_j(t')\rho(t')|m\rangle = \\
  &=\sum_{n,n',k}\delta_{mn'}\langle m'|Q_j|k\rangle \langle k|
  Q^{\dag}_i(\tau)|n\rangle \langle n| \rho(t')|n'\rangle
  e^{i\Omega^{m'n}_{mn'}t},
\end{align}
where $\Omega^{m'n}_{mn'}=(E_{m'}-E_{n}+E_{n'}-E_{m})$, $\tau=t-t'$,
and where the RDM elements have been rewritten as,
\begin{equation}
  \rho_{n',m}(t)=\sum_n\delta_{n,m}\rho_{n',n}(t)e^{i\omega_{n,m}t}.
\end{equation}

Terms 3,5,7 in the right side of \eqref{eq:rho_dot2} can be written in
a similar way.

The elements of the second term in \eqref{eq:rho_dot2} can be
rewritten as,
\begin{align}
  \nonumber &\langle m'|Q_i(t)\rho(t')Q^{\dag}_j(t')|m\rangle =\\
  &\sum_{n,n',k}\delta_{m,n'}\langle m'|Q_j|k\rangle \langle k|
  Q^{\dag}_i(\tau)|n\rangle \langle n| \rho(t')|n'\rangle
  e^{i\Omega^{m',n}_{m,n'}t},
\end{align}
and similarly for terms 2,6,8 in \eqref{eq:rho_dot2}.

The reservoir operator $F_{i}$ has no explicit time dependence; we can
therefore write, $\langle F^\dag_{\alpha}(t) F_{\beta}(t') \rangle =
\langle F^\dag_{\alpha} F_{\beta}(\tau) \rangle$, and $\langle
F^\dag_{\beta}(t') F_{\alpha}(t) \rangle = \langle
F^\dag_{\beta}(\tau) F_{\alpha} \rangle$.

Finally, for a given quantum dot state, we allow coupling with all
available reservoir states, while treating the reservoir energy as a
continuum.  That is,
\begin{equation}\label{eq:res_energy}
  \sum_{r}e_r \ \Rightarrow \
  N_R\left(\int^{\mu^R}_{\phi_{B}}de_r + \int_{\mu^R}^{\phi_{T}}de_r\right),
\end{equation}
where $e_r$ denotes the reservoir energies, $N_r$ and $\mu^R$ are the
2D density of states and chemical potential, respectively, for
reservoir $R$, and $\phi_{T}$ and $\phi_{B}$ are the top and bottom
energies of the band respectively.  With the redefinitions
\eqref{eq:system_correlation}-\eqref{eq:res_energy}, we rewrite
Eq.~\eqref{eq:rho_dot2} as,
\begin{widetext}
  \begin{align}\label{eq:rho_dot3}
    \nonumber \dot\rho_{m'm}(t)=\frac{1}{\hbar^2}\sum_{i,j}\int_0^tdt'
    \ \rho_{nn'}(t')&\left\{ \left[ Q^{j\dag}_{m'n}Q^i_{
          n'm}(\tau)\langle F^{\dag}_i(\tau)F_j\rangle+
        Q^j_{m'n}Q^{i\dag}_{n'm}(\tau)\langle
        F_i(\tau)F^{\dag}_j\rangle \right] \right.\\
    \nonumber &\ \ +\left[Q^{j\dag}_{n'm}Q^i_{m'n}(\tau)\langle
      F_jF^{\dag}_i(\tau)\rangle+Q^j_{n'm}Q^{i\dag}_{m'n}(\tau)\langle
      F^{\dag}_jF_i(\tau)\rangle\right]\\
    \nonumber \\
    \nonumber
    -&\sum_k\delta_{mn'}\left[Q^{j\dag}_{kn}Q^i_{m'k}(\tau)\langle
      F^{\dag}_i(\tau)F_j\rangle+
      Q^j_{kn}Q^{i\dag}_{m'k}(\tau)\langle
    F_i(\tau)F^{\dag}_j\rangle \right]\\
  \nonumber &\ \left. -\sum_k\delta_{m'n}
    \left[Q^{j\dag}_{n'k}Q^i_{km}(\tau)\langle
      F_jF^{\dag}_i(\tau)\rangle+ Q^j_{n'k}Q^{i\dag}_{km}(\tau)\langle
      F^{\dag}_jF_i(\tau)\rangle \right] \right\},\\
  \end{align}
\end{widetext}
where, $Q^i_{n'k}(\tau)\equiv \langle n'|Q_i(t-t')|k\rangle$.

Equation \eqref{eq:rho_dot3} is the generalized master equation for
the system, and describes the time evolution of the RDM.

By making use of Eq.~\eqref{eq:H_T_F} and \eqref{eq:res_energy}, we
derive the reservoir correlation functions appearing in
\eqref{eq:rho_dot3} to be
\begin{equation}\label{eq:correlations}	
  \begin{gathered}\langle F^{\dag}_{\alpha}(t)F_{\beta}\rangle 
    = K^r_{\alpha\beta} \Omega^{\alpha}_{\phi_B\mu^r}(t)  ; \
    \langle {F_{\beta} F_{\alpha} ^\dag(t)} \rangle 
    = K^r_{\alpha\beta} \Omega^{\alpha}_{\phi_T\mu^r}(t)\ \\ 
    \langle {F_{\beta}^\dag F_{\alpha}(t)} \rangle 
    = K^r_{\beta\alpha} \Omega^{\alpha \ *}_{\phi_B\mu^r}(t)  ; \ 
    \langle {F_{\alpha}(t) F_{\beta}^\dag} \rangle 
    = K^r_{\beta\alpha} \Omega^{\alpha \ *}_{\phi_T\mu^r}(t),
  \end{gathered}
\end{equation} 
where we have defined
\begin{equation*}
  \Omega^{\alpha}_{xy}(t) \equiv \frac{e^{i
      \omega_{x\alpha}t}-e^{i\omega_{y\alpha}t}}{t}, \ \ \ \  
  K^r_{\alpha\beta} \equiv \frac{i N_r}{\hbar}\left ( T^*_{\alpha}
    T_{\beta} \right )_r
\end{equation*}
representing the dynamics due to the interaction of the environment
with the quantum dot, where $\omega_{x\alpha}=(E_x - E_\alpha) /
\hbar$ is a frequency, $\left(T_{\alpha}\right)_r$ is the tunneling
coefficient appearing in Eq.~\eqref{eq:TunnelHamil}, $N_r$ is the
density of states for the 2DEG environment arising from the continuum
of reservoir states, and $t$ is time.

Comparing Eq.~\eqref{eq:rho_dot3} with Eq.~\eqref{eq:rho_dot_born} and
using Eq.~\eqref{eq:correlations} we finally arrive at the
microscopically derived expression for the memory kernel for the
present model

\begin{align}\label{eq:memory_kernel} 
  \nonumber R_{abcd}(t) =& \sum_{\alpha,\beta,r} K^r_{\alpha\beta}
  \left\{\Omega^{\alpha}_{\phi_B,\mu^r}(t) \Delta^{\alpha\beta}_{badc}
    - \Omega^{\alpha}_{\phi_T,\mu^r}(t) \Delta^{\alpha\beta}_{cdab}
  \right\}\\
  & + K^r_{\beta\alpha} \left\{ \Omega^{\alpha \ *}_{\phi_B,\mu^r}(t)
    \Delta^{\alpha\beta}_{abcd}- \Omega^{\alpha \ *}_{\phi_T,\mu^r}(t)
    \Delta^{\alpha\beta}_{dcba} \right\},
\end{align}
where
\begin{equation*}
  \label{eq:Delta}
  \Delta^{\alpha,\beta}_{abcd} \equiv \langle a|
  c^{\dag}_{\alpha}|c\rangle \langle d| c_{\beta}|b\rangle - \delta_{ac} \langle b| c_{\alpha} c^{\dag}_{\beta}|d\rangle 
\end{equation*}
represents the allowed system transitions for the present model, with
the indices $a,\ b,\ c,\ \text{and } d$ denoting many-body states;
$\alpha, \text{ and } \beta$ denote single-particle states, and $r \in
\{ source,\ drain\}$.  For the present analysis we assume a large band
limit where the top ($\phi_{T}$) and bottom ($\phi_{B}$) of the band
are taken to be effectively positive and negative infinity
respectively.

The dynamics of the density matrix can be calculated for a given set
of basis states, where Eq.~(\ref{eq:memory_kernel}) allows for
independent tuning and analysis of tunnelling rate anisotropies due to
asymmetries in the source and drain barrier widths, and due also to
asymmetries in the coupling between orbitals with differing angular
momentum.

In the following sections we treat the case where two transport
channels are available, and analyze the evolution of the system states
when the energy and interaction parameters are varied.

\section{The 2-Channel System}\label{sec:2_channels}

We present results for both the diagonal and off-diagonal elements of
the RDM, under the sequential transport model and formalism outlined in
the preceding sections.  For definiteness, we consider a quantum dot
with $N$ confined particles, and with two tunnelling transport
channels available within the bias window.  Each channel involves a
fluctuation in the particle-number between $N$ and $N \pm 1$, and we
choose them to involve either the ground or first excited state of the
$N$-particle system.  This is an experimentally relevant regime. (See,
for example refs.~[\onlinecite{0034-4885-64-6-201,chen:232109}])

In general, the availability of $k$ transport channels involves a
minimum of $2^k$ states.  We denote by $|0\rangle$ the
$(N-1)$-particle ground state of the system, $|1\rangle$ and
$|2\rangle$ denote ground-states of the $N$ and $(N+1)$-particle
system respectively, and $|3\rangle$ denotes the first excited state
of the $N$-particle system.  The four basis states of the system are
thus defined as,
\begin{align}
  \label{eq:states}
  \nonumber &|0\rangle = |(N-1)_{g.s.}\rangle,\ \ \  |1\rangle = |(N)_{g.s.}\rangle,\\
  &|2\rangle = |(N+1)_{g.s.}\rangle,\ \ \ |3\rangle =
  |(N)_{e.s.}\rangle.
\end{align}

\begin{figure}[tb!]
\begin{center}
  \includegraphics[width=20pc]{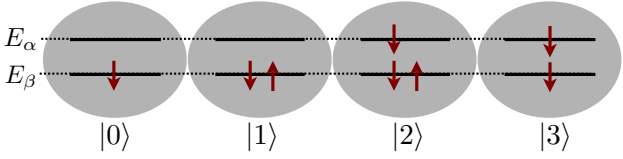}\hspace{2pc}%
  \begin{minipage}[b]{20pc}
    \caption{\label{fig:schem-repr-syst}(color online) Schematic representation of
      the system under consideration.  Four system states are
      considered depending on the relative occupation of the
      single-particle orbitals $|\alpha\rangle$ and $|\beta\rangle$,
      with energies $E_\alpha$ and $E_\beta$ respectively.  Each
      orbital can in principle have different tunnel couplings to the
      leads.}
  \end{minipage}
\end{center}
\end{figure}

A schematic representation of the system is shown in
Fig.~\ref{fig:schem-repr-syst}.

\subsection{Population Probabilities}\label{sec:pop_prob_2ch}

The occupation probabilities for the system to be in a given basis
state are obtained from the diagonal elements of the RDM.  We thus
derive the algebraic system Eq.~\eqref{eq:laplace_system1} for the
$2$-channel system with states defined by Eq.~\eqref{eq:states}.  The
diagonal elements in Laplace space are given by,
\begin{widetext}
  \begin{equation}
    \label{eq:laplace_rho_2ch} \tilde{\rho}_{nn}(s)=\sum_{i,j,k,l}\left[\frac{G^n_n\left(1-\frac{1}{2}G^i_jG^j_i-\frac{1}{3}G^i_iG^j_kG^k_i\right)+G^n_i\left(G^i_jG^j_j+G^i_jG^j_kG^k_k-G^k_jG^j_kG^i_i
        \right)}{1-\frac{1}{2}G^i_jG^j_i-\frac{1}{3}\left(G^i_jG^j_kG^k_i+G^i_jG^j_kG^k_lG^l_i\right)
        +\frac{1}{8}G^i_jG^j_iG^k_lG^l_k}\right]\Xi_{ijkl}
  \end{equation}
\end{widetext}
where $\Xi_{ijkl}$ is the symmetric permutation tensor given by,
\begin{equation}\label{eq:Xi}
  \Xi_{ijkl}=
  \begin{cases}
    1,&  \text{$ijkl$ permutation of $0123$}\\
0,&  \text{otherwise}
  \end{cases},
\end{equation}

and all indices run over the many-body system states defined in
Eq.~\eqref{eq:states}.

In Eq.~\eqref{eq:laplace_rho_2ch} the terms $G^i_i$ and $G^i_j$ are defined as
\begin{widetext}
\begin{subequations}\label{eq:Gs}
  \begin{equation*}
    \label{eq:Gii}
    G^i_i=D^{-1}\left[~\tilde{\rho}_{i,i}(0)+\frac{\tilde{\rho}^0_{1,3}\tilde{R}_{i,i,1,3}(s)}{s+i\omega_{1,3}-\tilde{R}_{1,3,1,3}(s+i\omega_{1,3})}+\frac{\tilde{\rho}^0_{3,1}\tilde{R}_{i,i,3,1}(s)}{s+i\omega_{3,1}-\tilde{R}_{3,1,3,1}(s+i\omega_{3,1})}\right],
  \end{equation*}
\begin{equation*}
    \label{eq:Gij}
    G^i_j=D^{-1}\left[~\tilde{R}_{i,i,j,j}(s)+\frac{\tilde{R}_{1,3,j,j}(s+i\omega_{1,3})\tilde{R}_{i,i,1,3}(s)}{s+i\omega_{1,3}-\tilde{R}_{1,3,1,3}(s+i\omega_{1,3})}+\frac{\tilde{R}_{3,1,j,j}(s+i\omega_{3,1})\tilde{R}_{i,i,3,1}(s)}{s+i\omega_{3,1}-\tilde{R}_{3,1,3,1}(s+i\omega_{3,1})}\right],
  \end{equation*}
with,
\begin{equation*}
  D= s-\tilde{R}_{i,i,i,i}(s)-\frac{\tilde{R}_{1,3,i,i}(s+i\omega_{1,3})\tilde{R}_{i,i,1,3}(s)}{s+i\omega_{1,3}-\tilde{R}_{1,3,1,3}(s+i\omega_{1,3})}+\frac{\tilde{R}_{3,1,i,i}(s+i\omega_{3,1})\tilde{R}_{i,i,3,1}(s)}{s+i\omega_{3,1}-\tilde{R}_{3,1,3,1}(s+i\omega_{3,1})}.
\end{equation*}
\end{subequations}
\end{widetext}

In the Markov limit (Eq.~\eqref{eq:rho_markov_2}) the system of
equations for the two-channel case is analytically solved in the time
domain to obtain,
\begin{equation}
  \label{eq:2channel_markov} \rho(t)_{nn}=\frac{\sum_{nijk}P_{nijk}+\sum_{ijkq}P_{ijkq}\rho_{qq}(0)e^{-\sum_{pq}W_{pq}t}}{\sum_{ijkl}P_{ijkl}}
\end{equation}
where, $P_{ijkl}=W_{ij}W_{jk}W_{kl}\Xi_{ijkl}$, with $\Xi_{ijkl}$
given by Eq.~(\ref{eq:Xi}), and the transition rate $W_{ij}$ given by,
\begin{align}\label{eq:W_0}
  \nonumber W_{ij}&= \frac{2\pi}{\hbar^2}\sum_{r}\int_{-\infty}^{\infty} de_\zeta  N_r |(T_\zeta)_r|^2\\
  \nonumber &\times \left[ \delta \left(e_\zeta-\omega_{ij}\right)\Theta \left( \mu^r-e_\zeta\right)+ \delta\left(\omega_{ij}-e_\zeta\right)\Theta \left(e_\zeta- \mu^r\right) \right].
\end{align}
where $r=\text{source},\text{drain}$, $\delta$ is the Dirac-delta
function, and $\Theta$ is the Heaviside function.

As a case study to validate the non-Markovian approach, we analyze the
effects of an orbital anisotropy on the dynamics of the system; we
define \emph{edge} and \emph{core} system orbitals by the strength of
their coupling to the reservoirs owing to the detailed shape of the
wave function.~\cite{ciorg02:collap.spin.singl} In 2-dimensional
parabolic confinement, for example, the particle's wave function is
composed of associated Laguerre polynomials where the core orbitals
are weakly coupled to the leads owing to a poor ($s$-type) overlap
with the reservoir states, whereas edge orbitals ($p$-type) are more
strongly coupled to the reservoirs since these wavefunctions penetrate
more deeply into the confining electrostatic barriers (see
illustration Fig.~\ref{fig:model}b).

In relation to Fig.~\ref{fig:schem-repr-syst}, we denote orbital
$|\alpha\rangle$ an edge orbital and $|\beta\rangle$ a core orbital.
Further denoting the orbital transmission by $T_{\alpha}$ and
$T_{\beta}$, we can define an orbital anisotropy parameter as
$\varepsilon=1-T_{\beta}/T_{\alpha}$, with $0< \varepsilon \le 1$.

The following energy parameters are used in the subsequent analysis:
applied bias $V_{\text{bias}} = 6$~meV symmetric about the Fermi
energy $\epsilon_{\text{Fermi}} = 30$~meV, with two channels within
the bias window, $E_{\alpha} = \epsilon_{\text{Fermi}} + 1$~meV and
$E_{\beta} = \epsilon_{\text{Fermi}} - 1$~meV.

The non-Markovian system of Eq.~\eqref{eq:laplace_rho_2ch}, is
evaluated numerically by performing an inverse Laplace transform as
described in Sec.~\ref{sec:evol-dens-matr}, thus obtaining results for
the diagonal elements of the RDM in the time domain.

\begin{figure}[tb!]
\begin{center}
  \includegraphics[width=20pc]{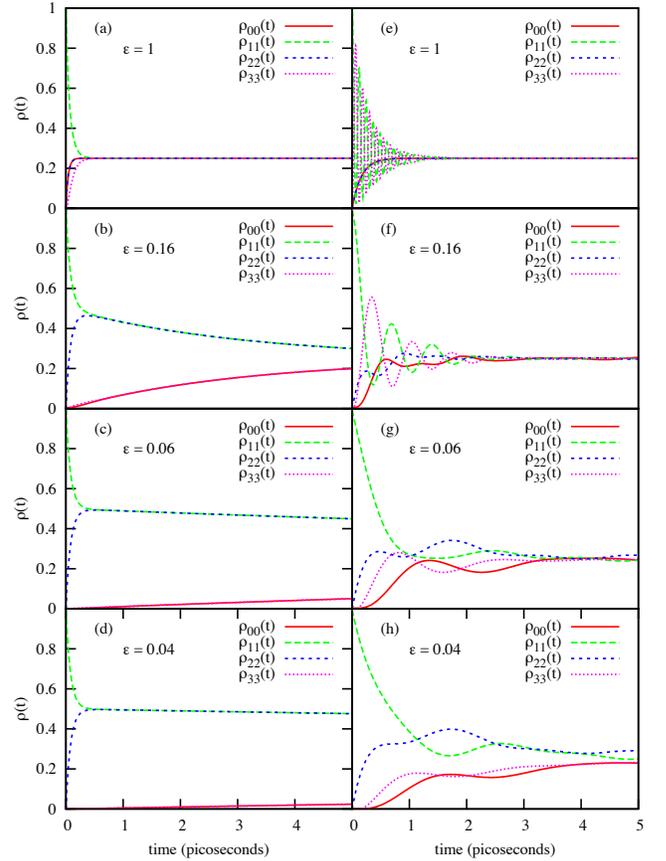}
  \caption{\label{fig:nonMarkov}(color online) Markovian and
    non-Markovian time evolution of population probabilities in a
    quantum dot with 2 transport channels and four states.  (See
    Fig.~(\ref{fig:schem-repr-syst}).)  The plots are for symmetric
    source and drain tunnel barriers, and varying orbital asymmetry.
    We assume a $6$~meV bias symmetric about a $30$~meV Fermi energy,
    and two transport channels at energies $\epsilon_{\text{Fermi}}
    \pm 1$~meV.  Plots (a) through (d) are results for the Markov
    limit (Eq.~(\ref{eq:rho_markov_2})), whereas plots (e) through (h)
    present results for the Non-Markovian theory.}
\end{center}
\end{figure}

The time evolution of the diagonal elements of the RDM in the large
bandwidth limit are shown for increasing orbital anisotropy in
Fig.~\ref{fig:nonMarkov}.  The set of figures on the left
(Fig.~\ref{fig:nonMarkov}.(a)-(d)) are the results obtained from the
Markov limit by solving
Eq.~\eqref{eq:rho_markov_2}, while the figures on the right
(Fig.~\ref{fig:nonMarkov}.(e)-(h)) present the non-Markovian results
based on Eq.~\eqref{eq:rho_dot_born}.  Four features in particular are
evident in the non-Markovian results which we now discuss.

First, in the non-Markovian results, we observe high-frequency
oscillations representing the coherent evolution of the occupancy
between states with the same particle number, $|1\rangle$ and
$|3\rangle$ (represented by RDM elements $\rho_{11}(t)$ and
$\rho_{33}(t)$ see Eq.~\eqref{eq:states}).  As expected these
oscillations are not present in the Markov limit.

Second, for small couplings to the core orbitals ($T_{\beta} \ll
T_{\alpha}$) the four probabilities couple into two distinct pairs,
states $|1\rangle$ and $|2\rangle$---the $N$ and $(N+1)$-particle
ground states---are strongly coupled, as are states $|0\rangle$ and
$|3\rangle$---the $(N-1)$-particle ground state and the $N$-particle
first-excited state.  We can understand why this occurs with recourse
to Fig.~\ref{fig:schem-repr-syst}.  The transition between states
$|1\rangle$ and $|2\rangle$ is through an edge orbital (with energy
$E_{\alpha}$) and this coupling is stronger than the transition
between $|1\rangle$ and $|0\rangle$ which involves a core orbital
(with energy $E_{\beta}$).  Similar arguments apply for the coupling
between states $|3\rangle$ and $|0\rangle$ (strong coupling) and
between $|3\rangle$ and $|2\rangle$ (weak coupling).  This pairing of
probability curves is also observed in the Markov limit, but again
without oscillatory behaviour.

Third, in the steady state ($t \rightarrow \infty$), all occupation
probabilities tend to the same value of 1/4.  This is seen regardless
of the tunnelling strengths of core and edge states as long as these
rates are non-zero.  The equal probability of 1/4 for each level can
be understood as being due to the symmetric barriers between the dot
and the source on the one hand, and between the dot and the drain on
the other.  Since these barriers are symmetric, any level will have an
equal probability of being either occupied or unoccupied.  Therefore
in the steady state and under symmetric barriers the levels will be,
on average, occupied and empty for the same amount of time.  This
result shows agreement with the Markov limit, where as expected all
dynamical states would share equal probability in the steady state
under symmetric barriers.

Fourth, although we observe that all four states are equally occupied
in the steady state, the time to \emph{actually reach} the steady
states does depend on the relative couplings to the core and edge
orbitals.  As expected, we observe a disagreement with the Markov
limit, where the Markovian results show that although the core or edge
states probabilities couple at a much shorter time, the overall steady
state is actually reached much later that in the non-Markovian
results.  

The fact that the time taken to reach the steady state increases as
the tunnelling to the core state decreases can be understood as the
effect of decreasing the available tunnelling pathways.  Fewer
pathways available means that the system takes longer to reach the
steady state.  This behaviour is exaggerated in the Markov
approximation.  In the limit of zero tunnelling to the core orbital,
for example, states $|3\rangle$ and $|0\rangle$ will \emph{never}
become occupied, and the remaining two levels will each reach an
occupation probability of $1/2$ rather than $1/4$, as expected by
preservation of the trace.  In this case the Markov limit agrees with
the non-Markovian results.

Although much of the recent work in the area of non-equilibrium
quantum dot systems has made use of Markovian-type
approximations,~\cite{Apalkov:2007p541,Imura:2007p227,
  pedersen:235314,Egorova:2003p212,Legel:2007p896,
  vaz_kyriakidis_2008} it is fundamental to acknowledge the strict
long-time limit requirement of these types of approximations in
general, and thus their unreliability on the transient behavior of the
system,~\cite{markov_exeption} especially when considering coherence
properties of the system at short time scales.  The disregard for the
memory of the system in effect neglects most of the details of the
coherent transient dynamics, and their use should only be invoked for
coarse grained phenomena.  Nonetheless, as expected, the Markov limit
does yield reasonable results at sufficiently long
times,~\cite{blum,Suarez:1992p3725} when the system has reached a
steady state, or when considering intermediate time scales of averaged
behavior, such as the average total current through a system.

\begin{figure}[tb!]
\begin{center}
  \includegraphics[width=18pc]{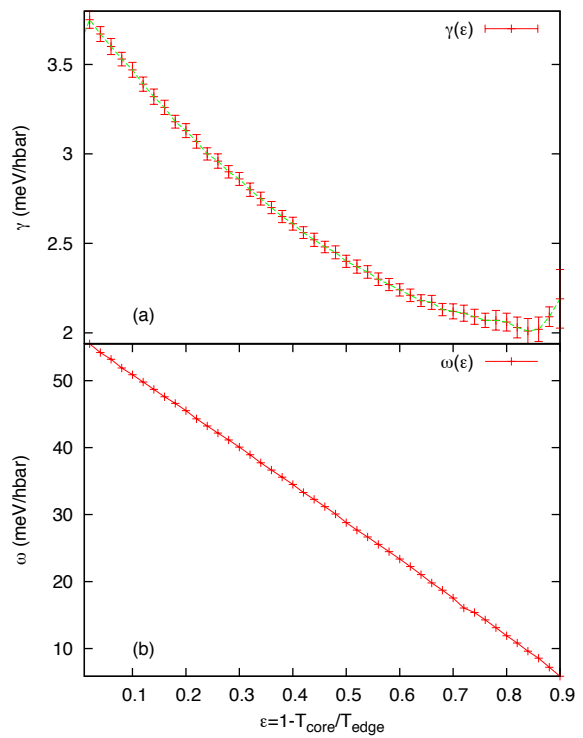}
  \caption{\label{fig:decay_diag}(color online) (a) Variation in the rate
    $\gamma$, of the decay of the oscillation of $\rho_{1,1}$. (b)
    Variation in the oscillation frequency of $\rho_{1,1}(t)$ as the
    orbital asymmetry is increased ($\varepsilon = 1$ is maximum
    asymmetry).}
\end{center}
\end{figure}

Focusing now on the transient results shown in
Fig.~\ref{fig:nonMarkov}(e--h), we have extracted the positions and
magnitudes of the peaks and fit~\cite{fit} these to a decaying
exponential of the form $ Ae^{-\gamma t}$. Results for the decay
constant $\gamma$ for each value of $\varepsilon=1-T_{\text{core}}/T_{\text{edge}}$ are shown in
Fig.~\ref{fig:decay_diag}(a).  We observe a slower decay rate as the
orbital asymmetry $(\varepsilon)$ is increased.
We interpret this result by considering the change in relative
occupation probability of the core channels \emph{relative to the edge
  channel} as tunnelling to and from them is increasingly suppressed.
In such a case the relative occupation of core states also decreases
and any contribution from these states to the coherent evolution of
the system will delay the overall decay to steady state, thus
increasing the decoherence time.

In a similar manner we have extracted the frequency of oscillation as
$\varepsilon$ is increased, where the results are shown in
Fig.~\ref{fig:decay_diag}. The linear decrease of the frequency with
an increase in $\varepsilon$ can be explained by again consider a
decrease in the core channel's occupation probability as $\varepsilon$
is increased.  Therefore, a coherent superposition between states
dependent on both core and edge channels will also exhibit the
relative decrease in probability from the core-orbital state.

In the following section, we look at the coherence elements of the RDM
under variations of the orbital anisotropy as well as of other relevant
coupling and energetic parameters.

\subsection{Coherence}
\label{sec:coherence}

The evolution of the quantum coherence between system states with the
same particle number,~\cite{fock_note} the Hilbert coherence, is
coupled to the evolution of the occupation probabilities through the
off-diagonal elements of the RDM.  For the present 2-channel case,
these coherence elements are derived in Laplace space to be,
\begin{equation}
  \label{eq:Laplaca_rho_offDiag} \tilde{\rho}_{13}(s+i\omega_{13})=\frac{\tilde{\rho}^0_{13}+\sum_{n=0}^3\tilde{\rho}_{nn}(s)\tilde{R}_{13nn}(s+i\omega_{13})}{s+i\omega_{13}-\tilde{R}_{1313}(s+i\omega_{13})}
\end{equation}
where the diagonal elements $\tilde{\rho}_{nn}(s)$ are given by
Eq.~\eqref{eq:laplace_rho_2ch}, and $\omega_{13}$ is the energy
difference between the $N$-particle states $|1\rangle$ and
$|3\rangle$.

\begin{figure}[b!]
\begin{center}
  \includegraphics[width=20pc]{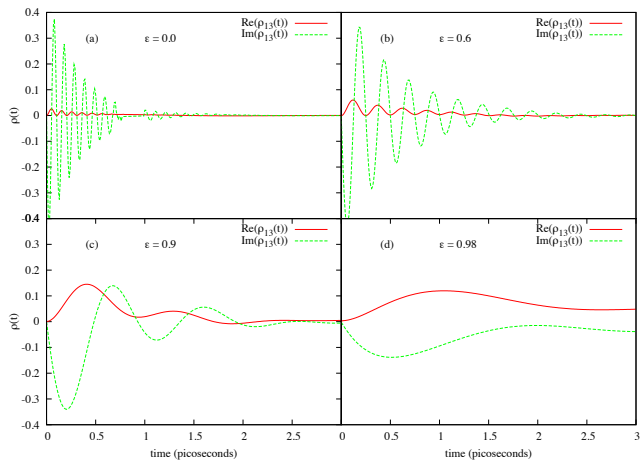}
  \caption{(color online) Time evolution of the real and imaginary
    parts of the off-diagonal elements of the RDM, $\rho_{13}(t)$,
    representing the Hilbert coherence of the system.  The plots are
    for symmetric source and drain tunnel barriers, and varying
    orbital asymmetry.  We assume a 6~meV bias symmetric about the
    Fermi energy, and two transport channels at energies
    $\epsilon_{\text{Fermi}} \pm 2$~meV.  The coupling strength for
    the edge channel is set at $\hbar T_{\alpha (\text{edge})}=0.5
    meV$, while the coupling to the core channel is varied.}
\label{fig:hilbert_coherence}
\end{center}
\end{figure}

We numerically evaluate the Bromwich integral of
Eq.~\eqref{eq:Laplaca_rho_offDiag} to obtain the real and imaginary
parts of the off diagonal element $\tilde{\rho}_{13}(t)=\langle
1|\tilde{\rho}(t)| 3 \rangle$ in time-space.  We are interested in the effects
of the barriers and energy configuration of the system on the
evolution of the coherence between dot states.  In this regard, the
important parameters considered are: 1) the orbital anisotropy, 2) the
source/drain barrier asymmetry, 3) the thickness of the barriers
(given inversely by the strength of the coupling $(T_{\alpha})_{r}$
between reservoir $r$ and dot orbital $\alpha$), 4) the energy level
spacing $\Delta E$ between the transport channels present within the
bias, 5) the applied bias voltage $V_{\text{bias}}$, and 6) the shift of energy
levels within the bias window.  The variations of these parameters are
discussed below.

\subsubsection{Orbital Anisotropy}

In Figure \ref{fig:hilbert_coherence}, we present results for the same
energetics and variation in the orbital anisotropy as shown for the
diagonal elements (Fig.~\ref{fig:nonMarkov}).  The Rabi-type coherent
oscillations observed in Sec.~\ref{sec:pop_prob_2ch} between the
diagonal elements $\rho_{11}(t)$ and $\rho_{33}(t)$
are also clearly seen in the evolution of $\rho_{13}(t)$ as expected. 

\begin{figure}[tb!]
\begin{center}
  \includegraphics[width=20pc]{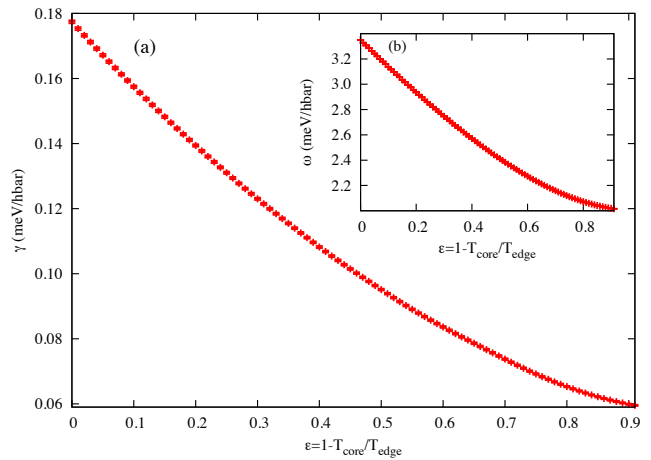}
  \caption{\label{fig:decay_diag_h}(color online) (a) Variation in the decay
    constant, $\gamma$, of the oscillation of $\rho_{13}(t)$, as the
    orbital asymmetry is increased ($\varepsilon = 1$ is maximum
    asymmetry). (b) Inset, variation in the oscillating frequency of
    $\rho_{13}(t)$.}
\end{center}
\end{figure}

Investigating the effects of this orbital anisotropy on the evolution
of the system, we focus on the decay constant and frequency of
oscillation of the off-diagonal elements of the RDM as
$(\varepsilon=1-T_{\text{core}}/T_{\text{edge}})$ is increased.  In
Fig.~\ref{fig:decay_diag_h}(a) the decay of the probability to a
steady state is fitted~\cite{fit} by an exponential function with rate
$\gamma$.  We observe a slower decay rate as the
orbital asymmetry factor $(\varepsilon)$ is increased.  As in
Sec.~\ref{sec:pop_prob_2ch}, we interpret this result by considering
the change in relative occupation probability of states dependent on
the core channel relative to states dependent on the edge channel, as
tunnelling to and from the core channel is increasingly suppressed.
In such a case the relative occupation of core states also decreases
and any contribution from these states to the coherent evolution of
the system will delay the overall decay to steady state, thus
increasing the decoherence time.

The effect of a variation of the orbital anisotropy on the frequency
of oscillation is shown in Fig.~\ref{fig:decay_diag_h}(b).  We again
observe a linear decrease of the frequency with an increase in
$\varepsilon$.  By the same argument as discussed in
Sec.~\ref{sec:pop_prob_2ch}, we consider a decrease in the occupation
probability of the core channel as $\epsilon$ is increased.  A
coherent superposition between states dependent on both the core and
edge channels will also exhibit the relative decrease in probability
from the core-dependent state.  Thus as expected, the coherent
evolution between states with the same particle number is described by
the off-diagonal element $\rho_{13}(t)$.

\subsubsection{Source/Drain Barrier Asymmetry}

\begin{figure}[tb!]
\begin{center}
  \includegraphics[width=20pc]{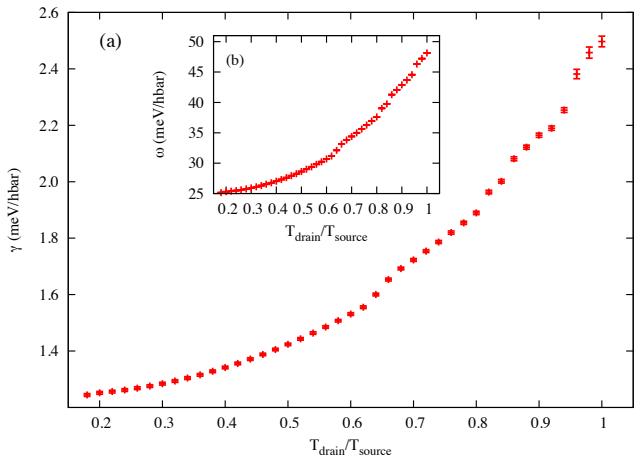}
  \caption{\label{fig:BA_decay_freq}(color online) (a) Variation in the decay
    constant, $\gamma$, of the oscillation of the coherence element
    $\rho_{13}(t)$, as the asymmetry between the source and drain
    barriers is increased ($T_{\text{drain}}/T_{\text{source}}\neq 1$). (b) Inset,
    variation in the oscillating frequency of $\rho_{13}(t)$.}
\end{center}
\end{figure}

Keeping now the orbital anisotropy fixed and varying the asymmetry
between the source and drain barriers, we observe that as the
thickness of one barrier relative to the other is arbitrarily changed,
there is a always an increase in the decay rate $\gamma$ (see
Fig.~\ref{fig:BA_decay_freq}a).  This can be understood by considering
the contribution to the decoherence rate due to various asymmetries:
In the case where the drain barrier is thicker than the source barrier, tunnelling
through the drain is suppressed relative to that through the source
barrier.  Therefore decoherence due to rapid tunnelling out of the dot
is also suppressed.  In the opposite case where the source barrier is
thicker, tunnelling into the dot is suppressed relative to tunnelling
out and thus, decoherence due to fast injection into the dot is also
suppressed.  Clearly, the decoherence will be greatest in the
completely symmetric case since the contributions from both electron
injection and extraction will be greatest.

Similarly, we find that the frequency of oscillation of the coherence
term ($\rho_{13}(t)$) is also maximal in the completely symmetric
case, and decreases otherwise (see Fig.~\ref{fig:BA_decay_freq}b).
This is expected since we can consider the frequency of oscillation to
be proportional to the energy splitting due to the coupling with the
reservoirs.  Note that the shape of the curve in both the decay
constant and frequency plots is quadratic with the relative thickness
of the barrier.  This is a consequence of the fact that, to leading
order, dissipative effects are quadratic in the system-bath couplings.
In the limit of infinitely thick barriers, there is no oscillation
since tunnelling though the barrier has been completely suppressed.

\subsubsection{Coupling Strength}

\begin{figure}[tb!]
\begin{center}
  \includegraphics[width=20pc]{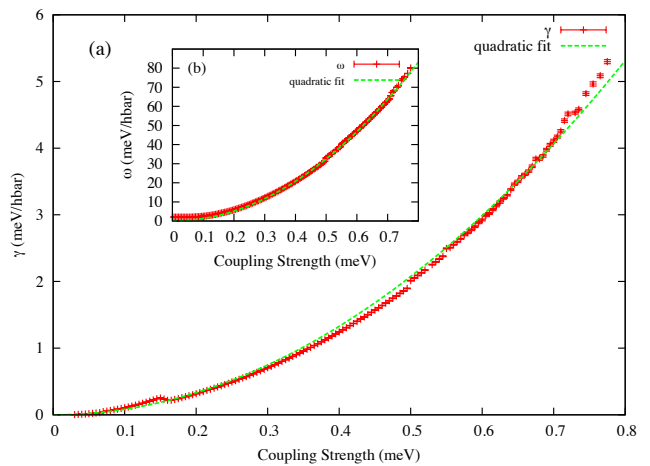}
  \caption{(color online) Variation of the decay $\gamma$ (a), and frequency of
    oscillation $\omega$ (b), as the overall coupling between orbitals
    and reservoirs is increased.  The overall coupling is changed by
    varying the strength of the tunnelling rate $T_{\text{edge}}$ and by
    keeping a fixed orbital anisotropy of
    $\varepsilon=1-T_{\text{core}}/T_{\text{edge}}=0.3$}
\label{fig:coupling_decay_freq}
\end{center}
\end{figure}

We now investigate the dependence of the coherence on the coupling
strength between the reservoirs and the available orbitals in the dot
while keeping the asymmetry between the barriers fixed.  For
simplicity, we also keep the ratio of the core-orbital tunneling rate
to the edge orbital tunneling rate fixed such that the coupling
strengths are varied at the same rate.

Figure~\ref{fig:coupling_decay_freq} presents the change in the decay
constant and the frequency of oscillation of the coherence element
$\rho_{13}(t)$ as the coupling strength is varied.  We observe that
both the decay constant and frequency closely follow a quadratic fit
on the coupling.  As mentioned in the discussion of the results of
barrier asymmetry, this is in agreement with the coupling effects
being quadratic to leading order.  We expect this to be a reasonable
result in cases where sequential transport is the dominant process.

\subsubsection{Energy Level Spacing}\label{sec:DeltaE}

\begin{figure}[tb!]
\begin{center}
  \includegraphics[width=20pc]{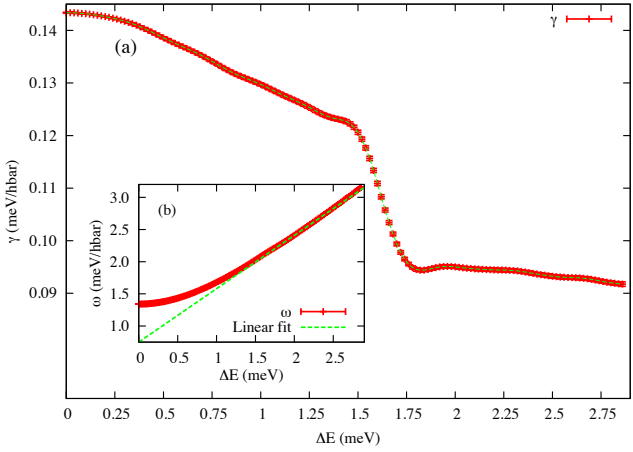}
  \caption{(color online) (a) Variation in the exponential decay
    constant $\gamma$, as the energy level spacing between the
    $|\alpha\rangle$ and $|\beta\rangle$ single particle levels is
    changed.  Error bars indicate how well the decay is approximated
    as being exponential.  (b) Variation in the frequency of coherent
    oscillation $\omega$ between many-body $N$-particle states
    $|1\rangle$ and $|3\rangle$.}
\label{fig:deltaE_decay_freq}
\end{center}
\end{figure}

Keeping now the barriers fixed, as well as the applied bias voltage,
we look at a variation in the energy level spacing $\Delta
E=E_{\alpha}-E_{\beta}$ between the $N$-particle states $|1\rangle$
and $|3\rangle$.  (All previous results considered $\Delta E = 2meV$.)
Results for the decay constant and frequency are shown in figure
\ref{fig:deltaE_decay_freq}(a) and (b), respectively.  The level
spacing is varied symmetrically about the Fermi energy.  In this case
we observe a decrease in the rate of decay as the level spacing is
increased (Fig.~\ref{fig:deltaE_decay_freq}a).  This result can be
understood in the following way: consider an electron in the source
reservoir which tunnels into one of the available levels in the dot,
and then tunnels out to the drain reservoir.  Since the dot states are
not eigenstates of the full Hamiltonian, the energy of the electron
within the dot is not well defined and can have a value within the
energy level spacing, $\Delta E$.  As the electron tunnels from the
source reservoir into the dot it can have at most an energy
$E_{\text{max}}=\epsilon_{\text{Fermi}}+1/2eV_{\text{bias}}$.  For the electron to then
tunnel into a drain state, it must have an energy of at least
$E_{\text{min}}=\epsilon_{\text{Fermi}}-1/2eV_{\text{bias}}+\Delta E$.  Therefore, the
available energy states of electrons that can tunnel into the dot from
the source reservoir is given by $E_{\text{max}}-E_{\text{min}}=eV_{\text{bias}}-\Delta E$.
Thus, the greater $\Delta E$ is, the fewer available energy states for
transport into the dot, and the longer lived the dot states are.  At
approximately $1.5\ meV$ we observe a sudden drop in the decay
constant leveling again after an increase in $\Delta E$ of ~$0.5\
meV$.  We investigate this feature by looking at its dependence on
other parameters, and find that as the coupling strength between the
reservoir and the dot is decreased, the feature decreases (see
Fig.~\ref{fig:deltaE_decay_freq}b).  We also observe that the slope of
the curve decreases after the drop, indicating a slower decrease of
the decay constant.

The dependence of frequency of oscillation of the coherence on the
energy level spacing is shown in Fig.~\ref{fig:deltaE_decay_freq}b.
The relevant energy differences giving rise to oscillations are the
level spacing $\Delta E$ on the one hand, and the level energy and the
edge of the bias window on the other.  For small $\Delta E$, it is the
latter which dominates and so $\omega$ shows weak dependence on
$\Delta E$ as shown in Fig.~\ref{fig:deltaE_decay_freq}b. For larger
$\Delta E$, it is $\Delta E$ itself which dominates and so we see
$\omega$ growing linearly with $\Delta E$.

\subsubsection{Applied Bias Voltage}\label{sec:Bias}

\begin{figure}[tb!]
\begin{center}
  \includegraphics[width=20pc]{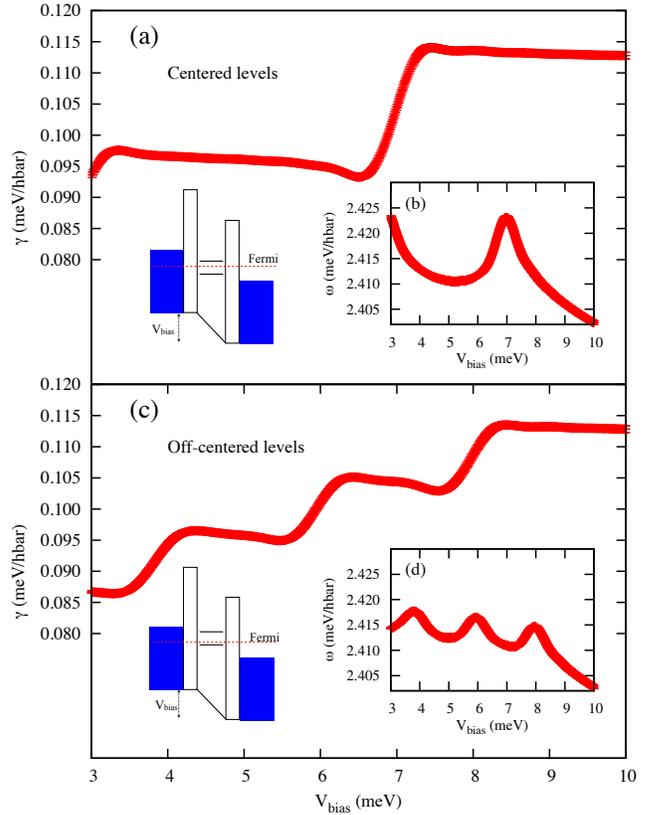}
  \caption{(color online) Variation of the decay constant $\gamma$
    (a,c) and of the frequency of oscillation $\omega$ (b,d) as the
    applied bias voltage is increased symmetrically about the
    transport channels.  In (a) and (b) the transport levels are kept
    symmetric about the center of the bias window, whereas in (c) and
    (d) the transport levels are shifted off-center within the bias
    window by $0.5 meV$. In all cases, $\Delta E=2 meV$.}
\label{fig:bias_decay_freq}
\end{center}
\end{figure}

A variation of the applied bias voltage has also been investigated,
with the results presented in Fig.~\ref{fig:bias_decay_freq}.  Here
$E_{\text{Fermi}}=30 meV, E_{\alpha}=31 meV, E_{\beta}=29 meV$.  For a
symmetric variation of the bias voltage about the Fermi energy, we
observe an overall increase of the decay constant as the bias voltage
is increased (Fig.~\ref{fig:bias_decay_freq}a).  We explain this
overall behaviour by considering that as the bias is increased, more
energy states are available in the reservoirs for electrons to tunnel
into or out of, electrons in the dot then have a larger number of
tunneling pathways, thus increasing relaxations, and decoherence.
This overall trend is also seen in the frequency of oscillation of the
coherence (Fig.~\ref{fig:bias_decay_freq}b) where, as the bias is
increased, the frequency generally decreases.  As discussed in
Sec.~\ref{sec:DeltaE}, the frequency of oscillation follows the
magnitude of the level spacing relative to the energy difference
between the levels and the edge of the bias.  Therefore, as the bias
increases relative to the level spacing, the frequency of oscillation
is expected to decrease.

Within this overall behaviour, we observe further structure in
Fig.~\ref{fig:bias_decay_freq} in the form of pronounced plateaus
occurring at specific intervals.  This structure can be explained by
noting that there are low frequency oscillatory envelopes (not shown)
on the evolution of the RDM elements given by the interplay of all
relevant energies in the system.  The intervals from one plateau to
the next are representative of the frequency of these slow envelope
oscillations.  The effect is also seen in the behaviour of the high
frequency oscillations, where as the bias in increased, these form
peaks with period given by the low frequency envelope
(Fig.~\ref{fig:bias_decay_freq}b).

We also find the number and position of these plateaus and peaks
depends on the position of the levels within the bias window.  Figures
\ref{fig:bias_decay_freq}c and \ref{fig:bias_decay_freq}d show an
increase in the number of plateaus as the levels inside the dot are
shifted off-center by $0.1 meV$ within the bias window
($E_{\alpha}=31.5 meV, E_{\beta}=29.5 meV$).  Therefore the low
frequency oscillations are strongly dependent on the relative energy
difference between the level spacing and the spacing between the
levels and the edge of the bias window.

Further analysis of these features will be the basis of future work by
the authors.

\subsubsection{Energy Level Shift Within Bias Window}

\begin{figure}[tb!]
\begin{center}
  \includegraphics[width=20pc]{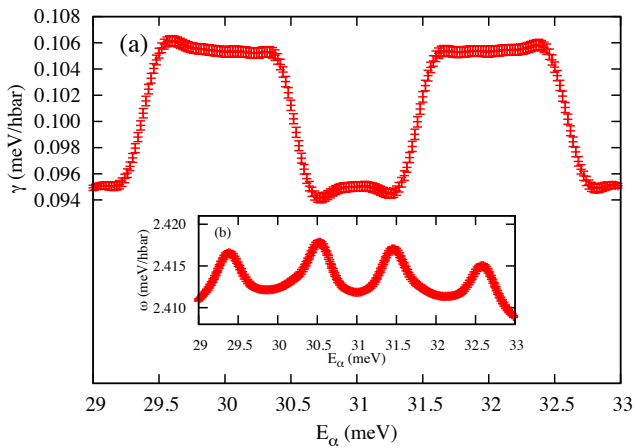}
  \caption{(color online) (a) Variation of the decay constant as the
    transport energy levels are shifted within the bias window. (b)
    Variation of the frequency of oscillation as the levels are
    shifted.  Here $\Delta E=2 meV$.}
\label{fig:AF_decayfreq}
\end{center}
\end{figure}

As mentioned in the discussion above, the position of the transport
channels relative to the center of the bias window has consequences on
the overall coherence between system states.  As before, we keep the
bias fixed at $6\ meV$ symmetric about a Fermi energy of $30\ meV$,
giving a chemical potential for the source reservoir of
$E_{\text{Fermi}}+\frac{1}{2}V_{\text{bias}}=33\ meV$.  Similarly the chemical
potential for the drain reservoir is then
$E_{\text{Fermi}}-\frac{1}{2}V_{\text{bias}}=27\ meV$.  By varying both levels
simultaneously and keeping the level spacing at $\Delta E=2\ meV$, we
can then move the edge level ($E_{\alpha}$) from $29\ meV$ to $33\
meV$ to sweep the levels over the entire window.  Figure
\ref{fig:AF_decayfreq} presents the result of this shift of the
position of $E_{\alpha}$.  In a similar fashion to the variation of
the applied bias, we observe structure in both the decay constant and
frequency of oscillation of the coherence of the system.  In
Fig.~\ref{fig:AF_decayfreq}a, the decay constant presents two plateaus
nearly symmetrical about the Fermi energy with size in the order of
$(0.5)\Delta E$.  For the frequency of oscillation
(Fig.~\ref{fig:AF_decayfreq}b) we see that it presents peaks at about
the same energies as the edges of the plateaus (similar to the results
of variation of $V_{\text{bias}}$ in Fig.~\ref{fig:bias_decay_freq}).  We
also observe an asymmetry of these peaks about the Fermi energy, where
the overall height decreases as the levels increase past the center of
the window.  As discussed in Sec.~\ref{sec:Bias}, the plateaus
appearing in the decay coefficient, and the peaks appearing in the
frequency of oscillation are due to envelope oscillations of low
frequency in the evolution of the RDM elements.

From these results, we can assert that the decoherence time is longest
when the levels are either centered in the bias window, or with at
least one level in resonance with the chemical potential of one of the
reservoirs.

Similar to the results of Sec.~\ref{sec:Bias}, we cannot definitively
ascribe a microscopic origin to these plateaus and peaks.  Analysis of
this structure will be the focus of future work.

\subsection{Summary}\label{sec:summary}

We now summarize the results of the variation of energy and coupling
parameters on the decoherence and frequency of oscillations of the
system.  For the sequential transport model presented, the following
conditions may independently increase the decoherence time in the
system: An increase in the difference of tunnelling rates between
available channels in the transport window, an increase in the barrier
asymmetry between source and drain barriers, a decrease of the overall
coupling strength between reservoirs and dot, an increase in the
energy level spacing between available transport channels, a resonant
case between available channels and reservoir Fermi energies or a
relatively small bias voltage, a symmetrical positioning of transport
levels within the bias window, or resonance between a transport level
and a reservoir Fermi energy.

Similarly, the frequency of oscillation can be decreased by: An
increase in the orbital anisotropy between available channels in the
transport window, an increase in the barrier asymmetry between source
and drain barriers, a decrease of the overall coupling strength
between reservoirs and dot, a decrease in the energy level spacing
between available transport channels, a resonant case between
available channels and reservoir Fermi energies, a relatively high
bias voltage, a symmetrical positioning of transport levels within the
bias window.  There is also the case that the frequency decreases in
certain intervals as the bias voltage is increased or as positions of
the energy levels are swept over the bias window. Table
\ref{table:summary} presents a tabular summary of the above results.

\begin{table}
\begin{tabular}{c c c}
\hline
\bf{Parameter} & \bf{Decoherence-time} & \bf{Frequency} \\ 
\hline
\hline
\bf{Level Spacing}      & \it{Increases} & \it{Increases} \\ 
\bf{Bias voltage}       & \it{Decreases} & \it{Decreases} \\ 
\bf{Off-center shift}   & \it{Decreases} & \it{Decreases} \\ 
\bf{Coupling strength}  & \it{Decreases} & \it{Increases} \\ 
\bf{Barrier asymmetry}  & \it{Increases} & \it{Decreases} \\ 
\bf{Orbital anisotropy} & \it{Increases} & \it{Decreases} \\
\hline
\end{tabular}
\caption{\label{table:summary} Effects of \emph{increasing} magnitude of
  parameters on coherence and frequency of
  oscillation.}
\end{table}

We thus find that an optimal lifetime for the coherence between
system states may be obtained by a suitable combination of the above
energetic and interaction regimes.  At the very least, by using the
above criteria, it may be possible to confirm the barrier and energy
parameters in experiment by studying decay rates and oscillatory
behaviour.

\vspace{12pt}

\section{Concluding Remarks}
\label{sec:conclusion}

We have developed a non-Markovian formalism for the transient
evolution of the reduced density matrix of a quantum dot weakly
coupled to source and drain reservoirs.  Within a tunneling
Hamiltonian approach, we have analytically derived an explicit memory
kernel describing the sequential transport dynamics of the system for
an arbitrary number of transport channels.  The results of the
analysis where compared with the Markov limit for a 2-channel
(4-state) system, where we verified marked differences in the
transient dynamics.  Apart from the expected absence of coherent
oscillatory behaviour in the Markovian results, it was also found that
the rate at which the system approaches a steady state differs
considerably between both theories in the regime of highly anisotropic
tunneling into distinct system orbitals.  It was found that the Markov
approximation shows significantly longer time to reach a steady state
when the tunnelling anisotropy is high, thus confirming its
applicability only in the long-time limit.  Through a comprehensive
and systematic analysis, the decoherence time and frequency of
oscillations observed in the non-Markovian results where found to
depend on both the energy parameters of the system as well as the
distinct coupling parameters to the reservoirs.  Finally, distinct
regimes have been outlined where both of these coherence
characteristics could be increased.

\section{Acknowledgements}

This work is supported by Lockheed Martin Corporation and by the
Natural Sciences and Engineering Research Council of Canada.

\bibliography{nonMarkov}

\begin{thebibliography}{45}
\bibitem[{rho()}]{rho_conditions}
\bibinfo{note}{This condition can be directly obtained from the normalization
  condition $Tr\{\rho(t)\}=1$, and the mixed-state condition $Tr\{\rho^2(t)\} <
  1$}.
\bibitem[{fit()}]{fit}
\bibinfo{note}{Our fit was performed using an implementation of the nonlinear
  least-squares Marquardt-Levenberg algorithm, where the error bars are
  calculated from the variance of the residuals.}
\end{thebibliography}
\bibliographystyle{apsrev}

\end{document}